\documentclass[12pt,preprint]{aastex}

\usepackage{epsfig}
\usepackage{rotating}
\usepackage{color}

\def\e10{\eta_{10}}
\def\Dtrans{\mbox{$D_{\rm trans}$}}
\def\simpropto{\lower.2ex\hbox{$\; \buildrel \propto \over \sim \;$}}
\def\ltsim{\lower.5ex\hbox{$\; \buildrel < \over \sim \;$}}
\def\gtsim{\lower.5ex\hbox{$\; \buildrel > \over \sim \;$}}

\def\etal{{\it et al.\ }}
\def\iso#1#2{\mbox{${}^{#2}{\rm #1}$}}

\def\b1#1{\iso{B}{1#1}}

\def\beq{\begin{equation}}
\def\eeq{\end{equation}}
\def\beqar{\begin{eqnarray}}
\def\eeqar{\end{eqnarray}}
\def\simlt{\lower.5ex\hbox{$\; \buildrel < \over \sim \;$}}
\def\simgt{\lower.5ex\hbox{$\; \buildrel > \over \sim \;$}}
\def\simpropto{\lower.2ex\hbox{$\; \buildrel \propto \over \sim \;$}}

\begin{document}

\title{The Impact of Star Formation and Gamma-Ray Burst Rates at High Redshift on Cosmic Chemical Evolution and Reionization}

\author{Elisabeth Vangioni}
\affil{Institut d'Astrophysique de Paris, UMR 7095 CNRS, University Pierre et Marie Curie, 98 bis Boulevard Arago,
Paris 75014, France}

\author{Keith~A.~Olive}
\affil{ William I. Fine Theoretical Physics
Institute, School of Physics and Astronomy, \\
University of Minnesota, Minneapolis, MN 55455 USA}

\author{Tanner Prestegard}
\affil{ School of Physics and Astronomy, \\
University of Minnesota, Minneapolis, MN 55455 USA}

\author{Joseph Silk}
\affil{
Institut d'Astrophysique, UMR 7095 CNRS, University Pierre et Marie Curie, 98 bis Boulevard Arago,
Paris 75014, France,
and\\ Department of Physics  and Astronomy, The Johns Hopkins University, Baltimore MD 21218 USA }

\author{Patrick Petitjean}
\affil{Institut d'Astrophysique de Paris, UMR 7095 CNRS, University Pierre et Marie Curie, 98 bis Boulevard Arago,
Paris 75014, France}

\and 
\author{Vuk Mandic}
\affil{ School of Physics and Astronomy, \\
University of Minnesota, Minneapolis, MN 55455 USA}

\begin{abstract}

\vskip-8.5in
\begin{flushright}
UMN-TH-3350/14 \\
TPI-MINN-14/25 \\
September 2014
\end{flushright}
\vskip+8in

Recent observations in the total luminosity density have led to 
significant progress in establishing the star formation rate (SFR) at high redshift.
Concurrently observed gamma-ray burst rates have also been used to extract
the SFR at high redshift.
The SFR in turn can be used to make a host of predictions concerning the
ionization history of the Universe, the chemical abundances, and supernova rates.
We compare the predictions made using a hierarchical model of cosmic chemical 
evolution based on three recently proposed SFRs: two based on extracting the SFR from
the observed gamma-ray burst rate at high redshift, and one based on the observed
galaxy luminosity function at high redshift. 
Using the WMAP/Planck data on the optical depth and epoch of reionization,
we find that only the SFR inferred from
gamma-ray burst data at high redshift suffices to allow a single mode
(in the initial mass function) of star formation which extends from z = 0 to redshifts $> 10$.
For the case of the more conservative SFR based on the observed galaxy luminosity function, the reionization history of the Universe 
requires a bimodal IMF which  includes at least a coeval  high (or intermediate) mass mode
of star formation at high redshift ($z> 10$).  Therefore, we also consider here a more general bimodal case which includes an early-forming 
high mass mode as a fourth model to test 
the chemical history of the Universe. We compute the abundances of several trace 
elements, as well as the expected supernova rates, the stellar mass density and the specific SFR, sSFR, as a function of redshift
 for each of the four models considered. We conclude that observational constraints on the global metallicity and optical depth
at high redshift favor unseen faint but active star forming galaxies as pointed out in many recent studies.

\clearpage

\end{abstract}


\section{Introduction}

Any model of galactic or cosmic chemical evolution
will depend on an assumed stellar initial mass function (IMF)
and a potentially measurable star formation rate (SFR).
In fact, only the convolution of the two is measurable through
the observed luminosity density.  Nevertheless, dramatic improvements
have been made in our understanding of the star formation history, particularly 
at high redshift. The pioneering work of \cite{Lilly}
using the Canada-France Redshift Survey started the path towards 
obtaining the comoving luminosity density at redshifts up to $z\sim 1$, quickly followed by
numerous studies which showed an intense period of star formation between redshifts
$1 < z < 2$ \citep{Madau96,Madau97,Connolly} and up to $z\lesssim 4$ \citep{Sawicki}
which showed some evidence for  a downturn in the inferred SFR. 

A large body of measurements of the luminosity density spanning redshifts up to $z = 6$
was compiled by \cite{Hopkins} and updated in \cite{HB06}.
These results point to a sharply rising SFR at low redshifts which 
peaks at around $z \sim 2$ and falls off at large redshifts. 
The redshift range was extended in a series of papers 
\citep{B07,B08,B11,Bouwens,O12,O13,O13-2,O14} which now include data out to 
a remarkable $z\approx 11$. The inferred fall off of the SFR, $\Psi$, at $z > 8$
appears to be quite steep, $\Psi \propto (1+z)^{-9}$ \citep{O13-2}.  

 \cite{leborgne09} had used the infrared (IR) galaxy counts to deduce the corresponding cosmic star formation history. Their study presents measurements of the IR luminosity function and they concluded that a sub-population of colder galaxies exists.
Recently, \cite{madau14} have reviewed in  great detail the cosmic star formation history and specifically discuss different complementary techniques which allow one to map the history of cosmic star formation.
Particularly important is the correction due to dust attenuation of the far-UV luminosity density in order to
be able to convert it to a SFR. One way of estimating the attenuation factor is a comparison
of the uncorrected inferred far-UV and far-IR SFRs \citep{burg}. This remains as one of the chief
uncertainties in establishing the SFR at high redshift.  Another issue is the contribution of emission lines to the apparent cosmic SFR density (CSFRD). However a general conclusion from the optical/UV/IR data  is that the stellar mass density  inferred from the CSFRD matches that observed over the entire observed redshift range \citep{madau14}.

Recently, in an attempt to reduce systematic uncertainties induced
due to the unknown stellar mass-to-halo mass relation, Markov Chain Monte Carlo
methods were employed on the large extant database up to $z = 8$ \citep{Behroozi}.
While there is certainly qualitative agreement (when normalized to the same IMF)
with the earlier work of \cite{HB06},
the new SFR of \citet{Behroozi} peaks at slightly lower redshift. 

It is well known that there may be significant uncertainties in extracting
the SFR from the luminosity function at large redshift. As noted above, this may be due to 
dust obscuration, but also more importantly, due to the fact that in any
scenario of hierarchical structure formation, early star formation
takes place in very faint galaxies (or proto-galaxies) and may be missed in 
existing surveys which are magnitude-limited. Indeed it has been argued \citep{Kistler13}
that integrating the luminosity function down to $M = -10$ (significantly below the lowest
measured value of $M_{vis} \simeq -18$) would lead to a significant increase in the SFR
at redshifts $z \gtrsim 4$. 
In addition, downsizing, which is both observed, (e.g. \citet{Behroozi, 2014MNRAS.439.1459Z}), and theoretically interpreted via the impact of 
feedback models on gas consumption \citep{2014arXiv1405.3749G, 2014arXiv1407.7040S}, further complicates matters.

An alternative to searching for star formation in low luminosity galaxies is
possible if we can connect the rate of gamma-ray bursts (GRBs) at high redshift
to the SFR \citep{totani1997a,wijers1998a,mao1998a,porciani2001a,bromm2002a,chary2007a}.
Recent attempts at making such a connection \citep{Y08,kistler2008a,Kistler09,wd09,wyithe2010a,RE,wangetal,Kistler13,wang}
have all indicated a higher SFR at high $z$ relative to methods using luminosity functions.
Using GRBs to trace the SFR, one would infer a much slower fall-off as a function of 
redshift, $\Psi \propto (1+z)^{-3}$ \citep{RE,Kistler13,wang} for $z> 4$. 
We note that the relation between the SFR and the GRB rates is very uncertain and the
method to derive the SFR redshift evolution from GRBs includes numerous
assumptions and still largely unknown biases (Vergani, private communication, see also \citet{Vergani13, Vergani14}). 
However, this is still a promising method and we will use these results as illustrative cases.
Noting the systematic uncertainty in connecting
the rate of GRBs with the SFR, attempts at modeling the relation
have produced somewhat more moderate results for the SFR at high $z$ \citep{wp,JP,Trenti,sok}, still 
consistent with the lower bounds found in \citet{Kistler13}.

The choice of SFR has direct consequences on the chemical and reionization history 
of the Universe. A steeply falling SFR, as obtained from the luminosity density
at high redshift (cf. \cite{Behroozi,O13-2,O14}), is likely to be insufficient for explaining
the optical depth extracted from cosmic microwave background (CMB) data \citep{wmap,planck}
and an additional high redshift mode of star formation which can be associated with a top-heavy
IMF is one way of alleviating this problem \citep{daigne1,daigne2,rollinde}. Moreover, the early enrichment of the IGM
similarly points to an additional massive mode beyond that extracted from the luminosity function
\citep{daigne1,daigne2,rollinde,shull12,trenti12}. 

In contrast, it has been argued that reionization \citep{gallerani,wang} and metal enrichment
\citep{wangetal} may better match observations if instead the SFR implied from the GRB rate is used.
This conclusion is especially apparent in the comparison of DLA data \citep{rafelski12} and cluster data as compiled by \cite{madau14}, higher ionization states resulting however in smaller inferred metallicities. We note also that the specific SFR (sSFR) is in better agreement  with that inferred from the cosmic star formation rate density  over the redshift range from 0 to 8 \citep{madau14}, if the emission line-corrected values of \cite{stark13} are used.

Here, we will explore the consequences of three choices for the SFR:
1) The SFR derived from the luminosity function of faint high redshift galaxies
from \citet{Behroozi} supplemented with the high redshift observations of \citet{B11,O12,O13,O13-2, O14}; 
2) The SFR implied by the Swift GRB rate from \cite{Kistler13};
3) An intermediate case where the SFR is scaled down by 0.3 dex, 
based on the models of \citet{Trenti,bs}.
For the normal mode of star formation which extends to the present day,
we will assume a Salpeter IMF. When appropriate, we will assume 
a high redshift mode of star formation which peaks at a redshift $z \gtrsim 10$.
While it is generally assumed that the initial mode of star formation is massive
\citep{bromm,bromm2}, it is possible that the high $z$ mode is dominated by
intermediate mass stars \citep{yoshii,smith09,schneider10,safra10}.
In all cases, we will work in the context of a hierarchical model for structure formation. 
This is coupled to a detailed model for cosmic chemical
evolution \citep{daigne1,daigne2,rollinde}, and allows us to 
keep track of the ionization history of the Universe, track the abundances  
of many of the elements produced in massive and intermediate mass stars as a function of redshift,
as well as track the rate of supernovae of type II (SNII), the sSFR,  and the stellar mass density.

In the next section, we briefly describe the model of cosmic chemical evolution
employed and our parameterization of the four choices of SFRs\ based on the \citet{sp03} 
form for the SFR. For the case where an additional 
mode of star formation is required, we optimize the choice of the SFR by 
scanning the parameter space and minimizing a $\chi^2$ likelihood function. 
We then compute the resulting ionization and chemical history 
of the Universe for each of the SFRs considered.  A priori, we assume
only a normal mode of star formation which extends to the present day.
When this is found to be insufficient, we  add  a complementary high $z$ mode
in order to achieve concordance for both the reionization and metal enrichment.
The resulting chemical evolution and a comparison of the four SFRs is  given in section 3. 
A discussion of our results is given in section 4.

\section{The SFR at High Redshift}
Our work here is developed from  a model of hierarchical structure formation 
based on the Press-Schechter formalism \citep{ps} which determines the rate at which 
structures accrete mass. The model includes exchanges of baryonic mass between
the gas within (the ISM) and exterior to (the IGM) structures. We assume a minimum 
mass for star-forming structures of $10^7 \rm M_\odot$.  Details of the model can be found in 
\citet{daigne1,daigne2,rollinde}.

In all cases, we assume an IMF with a single Salpeter slope ($x=1.35$). 
The normal mode of star formation includes stellar masses between
0.1 M$_\odot$   and 100 M$_\odot$. When this is supplemented with
a high redshift mode, we assume a mass range of 36 M$_\odot$   to 100 M$_\odot$
for high mass stars.
Our SFR is always parameterized using the form given by \citet{sp03}
\begin{equation}
\psi(z) = \nu\frac{a\exp(b\,(z-z_m))}{a-b+b\exp(a\,(z-z_m))}\,.
\label{shsfr}
\end{equation}
The amplitude (astration rate) and the redshift
 of the SFR maximum are given by $\nu$ and  $z_m$ respectively, while
$b$ and $b-a$ are related to its slope at low and high redshifts
respectively. 

To determine the ionization history, we take the 
evolution of the volume-filling fraction of ionized regions to be:
\beq
\frac{\mbox{d}Q_{{\rm ion}}(z)}{\mbox{d}z} = \frac{1}{n_{\rm
 b}}\frac{\mbox{d}n_{{\rm ion}}(z)}{\mbox{d}z}-\alpha_{{\rm B}}n_{{\rm b}}C(z)
   Q_{{\rm ion}}^{2}(z)\left(1+z\right)^{3}\left|\frac{\mbox{d}t}{\mbox{d}z}\right|\mbox{\ ,}
\eeq
where $n_{\rm b}$ is the comoving density in baryons, $n_{{\rm ion}}(z)$ the comoving density
of ionizing photons,  $\alpha_{{\rm B}}$ the recombination coefficient,  and $C(z)$ the
clumping factor. This factor is  taken from \citet{greif06} and  varies from a value of 2 at $z\leq20$ to a constant value of 10 for 
$z<6$. $dt/dz$ is taken to be the standard form for a $\Lambda$CDM cosmology
with a density of matter $\Omega_\mathrm{m}=0.27$ and a density of ``dark energy'' $\Omega_\mathrm{\Lambda}=0.73$ and taking $H_{0}=71\ \mathrm{km/s/Mpc}$. The escape fraction,  $f_{\rm esc}$, is set to 0.2 for each of our assumed modes of star formation . The number of ionizing photons  for massive
stars is calculated using the tables given in \cite{schaerer02}. Finally, the Thomson optical
depth is computed as in \cite{greif06}:
\begin{equation}
\tau =c\sigma_{{\rm T}}n_{\rm b} \int_{0}^{z}dz'\,Q_{{\rm ion}}(z')\left(1+z'\right)^{3}\left|\frac{\mbox{d}t}{\mbox{d}z'}\right|\mbox{\ ,}
\end{equation}
where $z$ is the redshift of emission, and $\sigma_{{\rm T}}$ the Thomson scattering
 cross-section. The latest result for the optical depth from the \citet{planck} is based on the nine year \citep{wmap} polarization data  and the two studies of optical depth are  not substantially different.  Here we use the \citet{wmap} result of $\tau = 0.089 
\pm 0.014$ to compare with model predictions.

In what follows we consider four choices for the SFR. We begin with the 
SFR inferred from the GRB rate \citep{Kistler13} which has the highest 
SFR at high $z$ of the models considered (model 1). \citet{Trenti}  argued that the normalization used to 
construct the SFR from the GRB rate should not evolve beyond $z=4$ because most galaxies 
would have low enough metallicity that the GRB rate saturates. In fact the specific mass accretion rate from simulations can be applied to compute the star formation rate  by assuming the empirical ratio of stellar to halo mass obtained by
normalizing to  data at $z\simlt 4,$ to successfully account for the observations to $z\sim 8,$  and then to predict the specific SFR (sSFR - discussed below) and SFR, to $z\sim 15$
 \citep{bs}. This corresponds to our second model below (model 2). 
 The third model considered is based on the galaxy luminosity function \citep{Behroozi} (model 3). 
We will show, however, that this model marginally fails, for reasonable escape fractions, to give sufficient CMB optical depth, and develop a variant (model 4) in which a complementary, high $z\simgt 10$, mode of massive star formation is added to simultaneously account for enrichment and ionization.

\subsection{The SFR based on the GRB rate}
\label{kist}

\begin{figure}[htb!]
\begin{center}
\epsfig{file=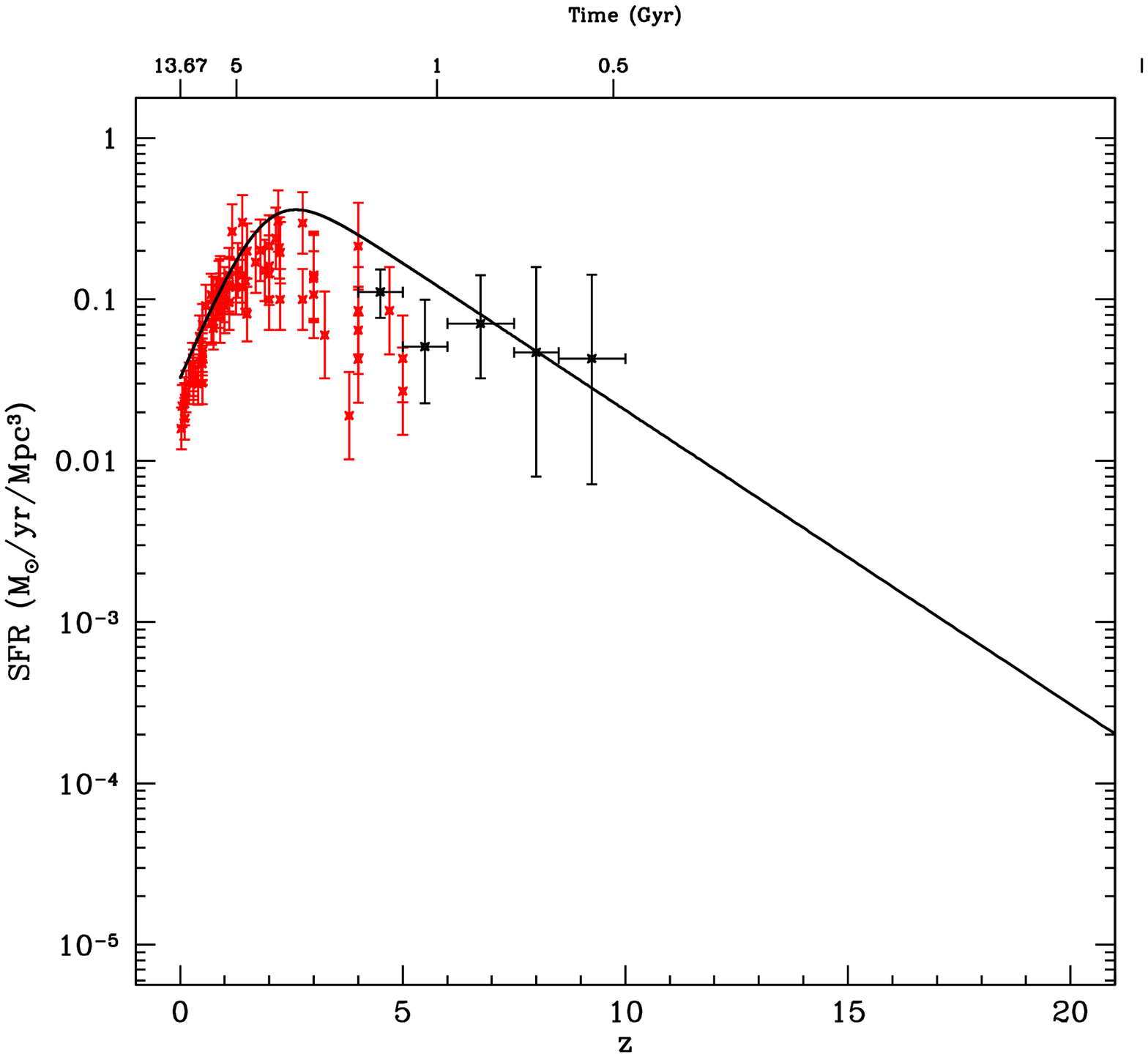, height=3in}
\epsfig{file=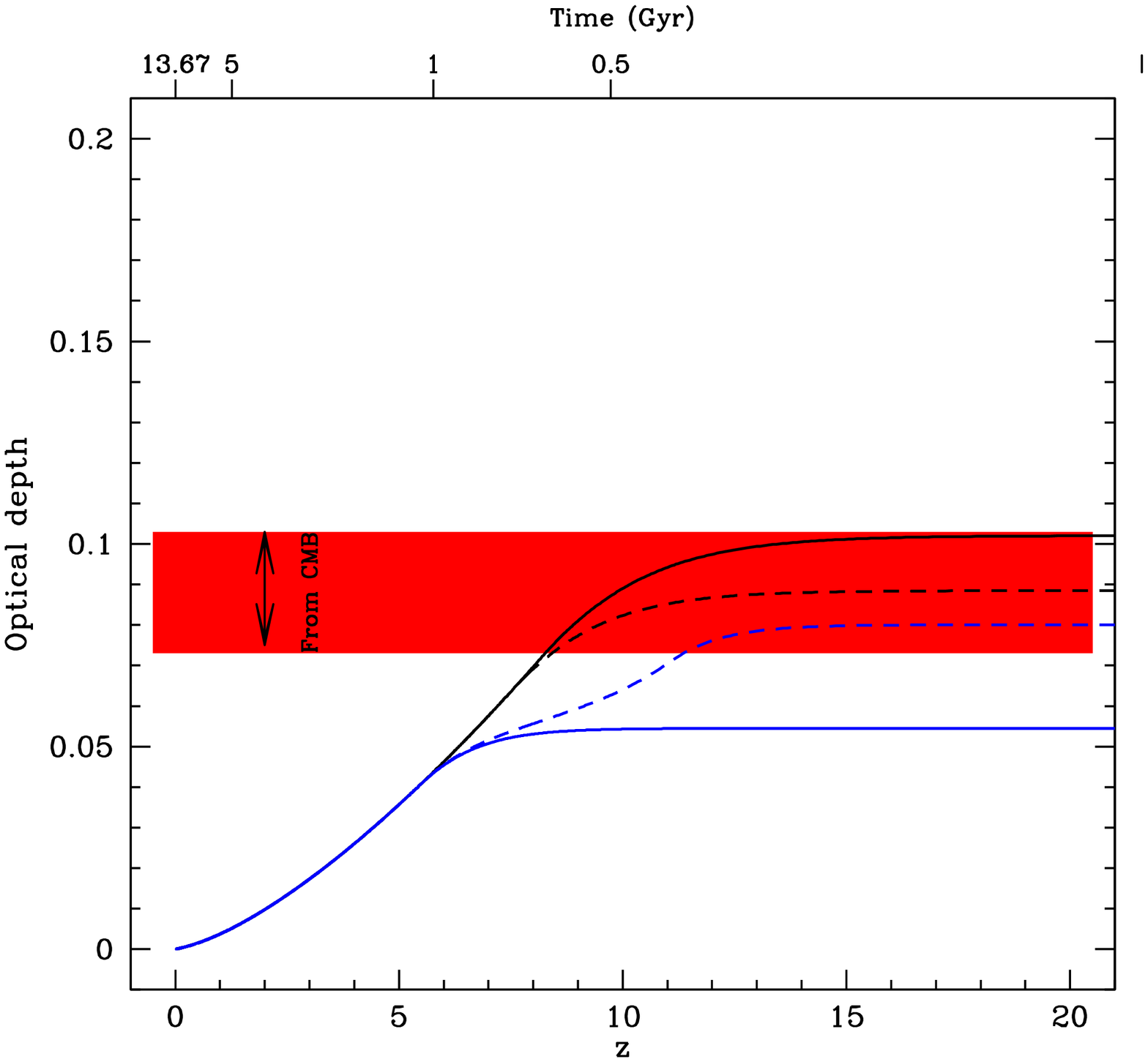, height=3in}
\vskip -.2in
\epsfig{file=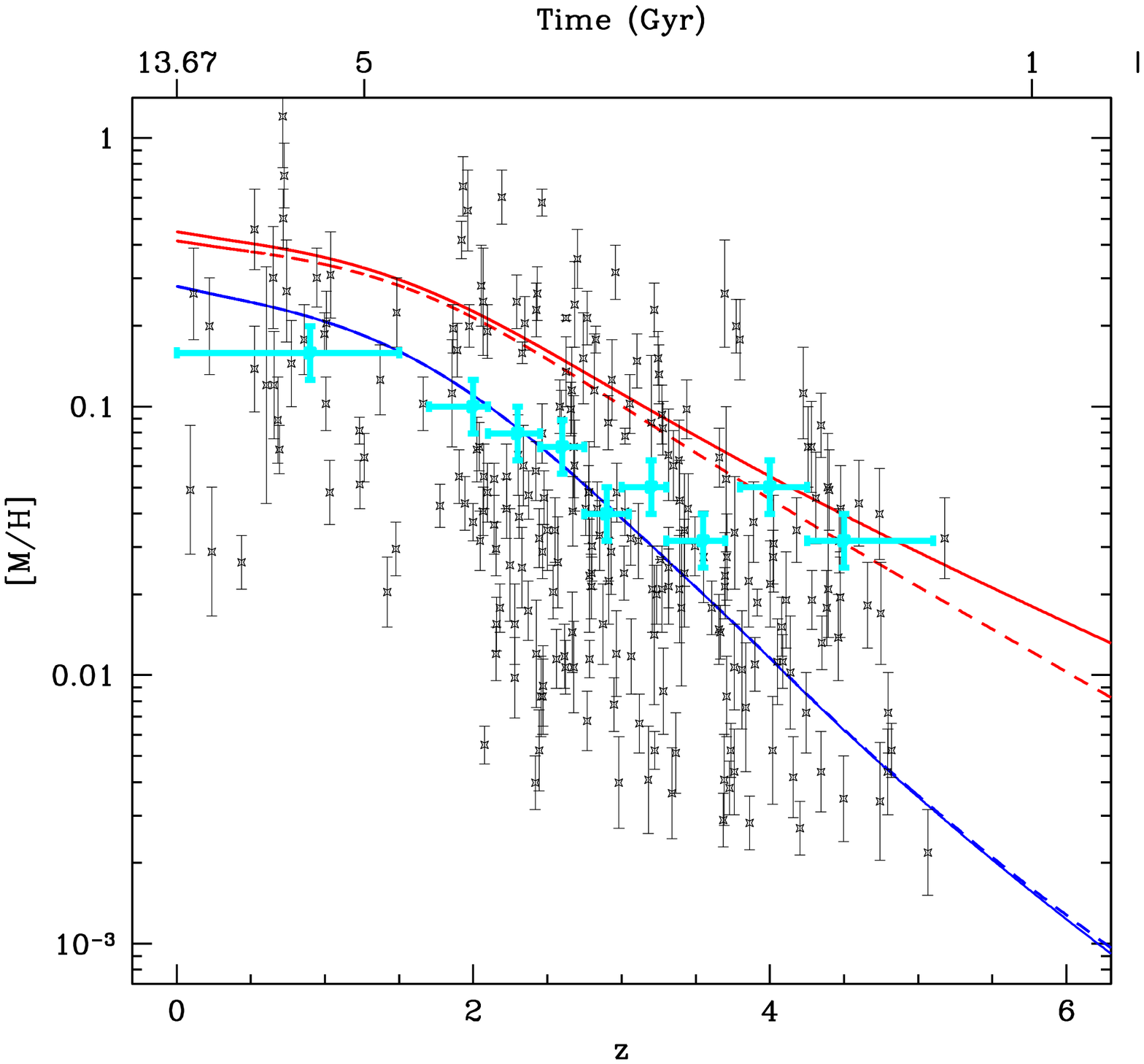, height=3in}
\epsfig{file=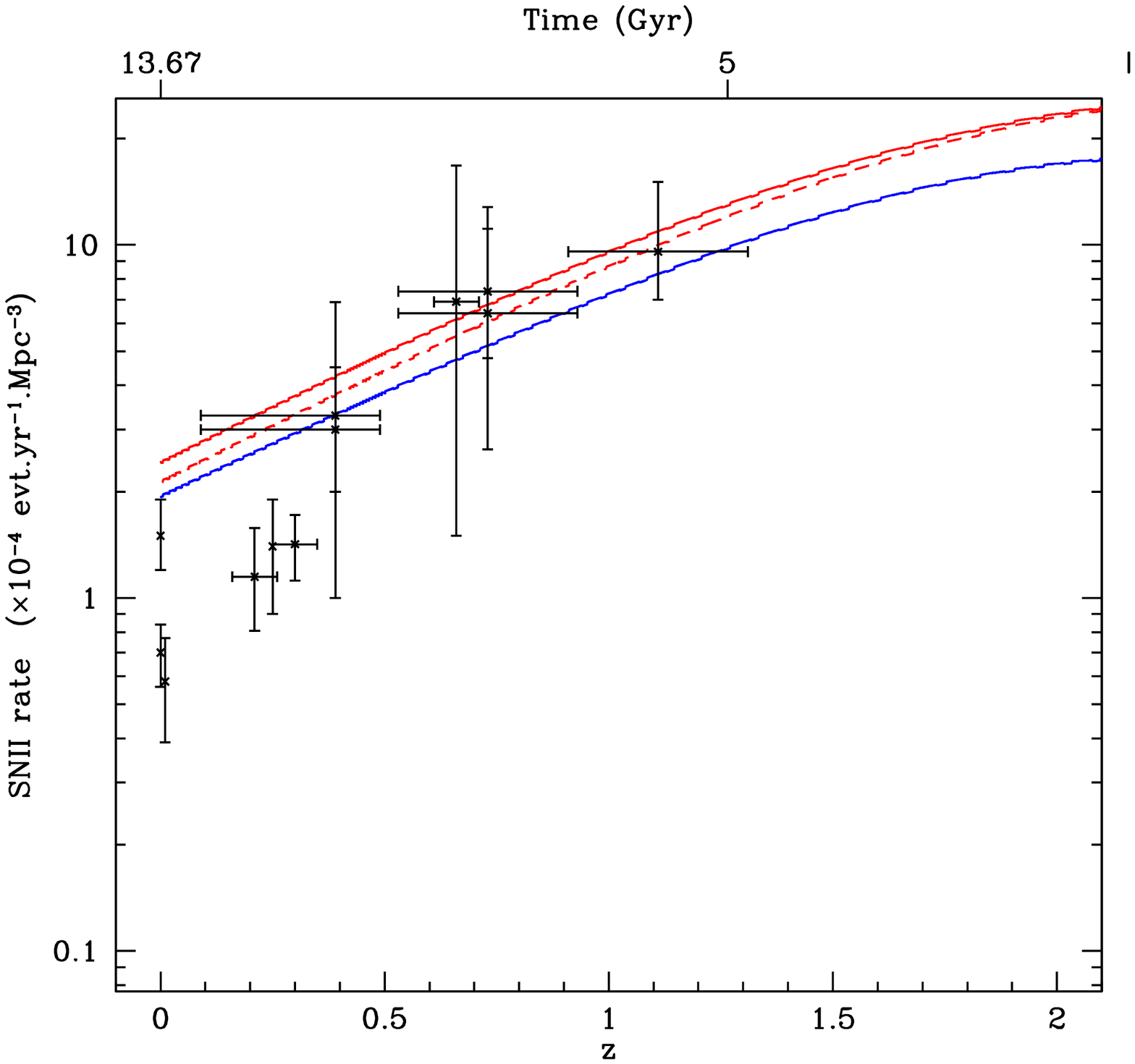, height=3in}
\end{center}
\caption{ a) Fit to the SFR based on the GRB rate as derived in \citet{Kistler13} (model 1); b) the derived optical depth for models 1 (solid black curve), 2 (dashed black curve), 3 (solid blue curve), and 4 (dashed blue curve); c) the (log of the) metallicity abundance relative to the solar metallicity for models
1 (solid red curve), 2 (dashed red curve), 3 (solid blue curve), and 4 (dashed blue curve); d) the derived SNII rate using the same color coding as in panel c).
All quantities are shown as functions of redshift. Observational constraints are given in the text.
Note that the SFR for models 2, 3, and 4 are presented separately in Figs.~\ref{fig:k2}, \ref{fig:b1}, and \ref{fig:b2}
for clarity as we have used different observational constraints for the SFR in each case.
\label{fig:k1}
}
\end{figure}

Keeping in mind the uncertainties in converting the GRB rate to 
a global SFR,
we begin by first considering the SFR obtained from the GRB rate in
\citet{Kistler13}. We have fit the SFR to the  \citet{sp03} form 
and find $\nu = 0.36$  M$_{\odot}$/yr/Mpc$^{3}$, $z_m = 2.6, a = 1.92$, and $b = 1.5$.
The assumed SFR is shown in the upper left panel of Fig.~\ref{fig:k1}a
and is chosen to fit the data from \citet{Kistler13} represented by the five black points 
(with high SFR and large redshift). The red points providing the bulk of the data
at low redshift are taken from \citet{Behroozi}\footnote{These points have been scaled upward
by a factor of 1.7 relative to the values given in \citet{Behroozi} to account for 
our choice of a Salpeter IMF (Behroozi, private communication).   
In \citet{Behroozi}, the \cite{chab} IMF was chosen.}.
The slope at low $z$ (below the peak), is determined by data compiled in \citet{Behroozi},
and as one can see, this choice of the slope parameters fits the high redshift data of 
\cite{Kistler13} quite well. The optical depth for this model as a function of redshift in shown 
by the solid black curve in the upper
right panel of Fig.~\ref{fig:k1}b and falls (barely) within the 68\% CL limit of the WMAP result for the 
optical depth, $\tau$. For this model, $\tau(z=30) = 0.102$ and $z_I = 9.84$ where 
$z_I$ is defined as the redshift at which $Q_{\rm ion} = 0.5$.
The WMAP value for $z_I$ is 11.1 $\pm$ 1.1.
 Also shown in Fig.~\ref{fig:k1} are the overall metallicity relative 
to the solar metallicity (1c, solid red curve in the lower left)) and 
SNII rate (1d, solid red curve in the lower right) as functions of redshift.

We note that had we fit the \cite{Behroozi} and \cite{Kistler13} data alone,
a better fit to the form of the SFR we are using would be 
$\nu = 0.16$  M$_{\odot}$/yr/Mpc$^{3}$, $z_m = 1.9, a = 2.76$, and $b = 2.56$,
resulting in a much flatter fall off at large $z$.  However in this case, we find an optical depth 
$\tau \approx 0.14$ greatly in excess of the WMAP result. The SFR shown in Fig.~\ref{fig:k1}a is
about as flat as one can allow while remaining within 1$\sigma$ of the WMAP determination of $\tau$.

In Fig.~\ref{fig:k1}c,  observational data (black points) come from \citet{rafelski12}.
There, they present chemical metallicity measurements [M/H] (coming from different elements: Si, S, Zn, Fe, O)  for 47 damped Ly$\alpha$ (DLA) systems, 30 at $z > 4$, observed with
the Echellette Spectrograph and Imager and the High Resolution Echelle Spectrometer on the Keck telescopes. They combine these
metallicity measurements with 195 from previous surveys, which were drawn from the SDSS-DR3 and SDSS-DR5 surveys for DLA systems performed by \cite{prochaska05} and  \cite{prochaska09} respectively. Moreover, we have added the nine  mean points (cyan points) from \cite{rafelski12} (see their figure 11) 
where horizontal error bars are determined such that there are equal numbers of data points per redshift bin. We note that at high redshift ($z >4$), models 1 and 2 (red lines) are a good fit to these mean Z observations. This could support the idea that GRBS could be better metal tracers (and also SFR tracers) at high z.

The GRB rate-inferred cosmic star formation history is anti-biased with regard to the dark matter distribution according to \cite{JP}, who argue that the implied preference for regions of high star formation accounts for the apparent enhancement in the corresponding star formation history. This conclusion is supported by our chemical evolution study: we obtain a slope that matches the data for the GRB-inferred cosmic star formation history, and agrees in normalization if we reduce  
the effective yield by a factor $\sim 3.$

In Fig.~\ref{fig:k1}d, data points indicating the cosmic SNII rate are available from several observational surveys \citep{li09, dahlen12, botticella08, graur11, bazin09, melinder11}. Recently, the estimate for the  local rate (now 1.5 $ \times10^{-4}$ events/yr/Mpc$^{3})$ has been increased. \citet{mattila12} argued that supernovae have been missed in optical surveys due to dust obscuration. The new higher rate at $z = 0$, is now in much better agreement with the types of models
we are considering.

Both quantities (the total metallicity and SN rate) are sensitive to the assumed model of chemical
evolution and can be used to discriminate between models and make reasonable parameter choices.
Like the optical depth, in this case, both the metallicity and SN rate are somewhat high though acceptable. 
Clearly a single mode of star formation is sufficient to explain these data.

\begin{figure}[htb!]
\begin{center}
\epsfig{file=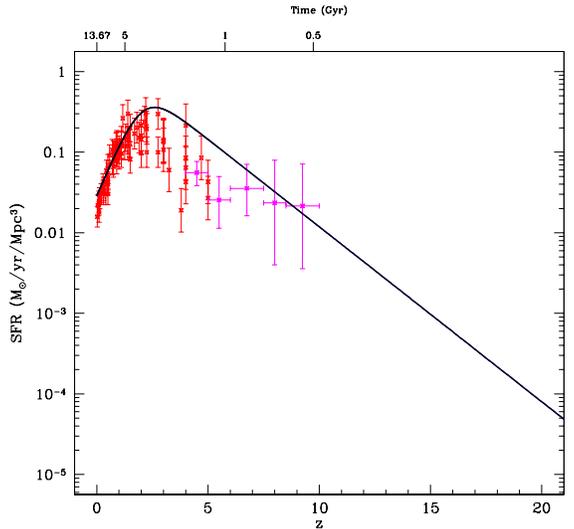, height=3in}
\end{center}
\caption{ As in the upper left panel of Fig.~\ref{fig:k1}a the SFR for model 2, using however, the normalization derived in \citet{Behroozi} for the five points derived from the GRB rate, now shown in magenta.
\label{fig:k2}
}
\end{figure}

The above choice for the SFR can be considered as an upper limit. 
It was derived using a normalization of the GRB rate to the SFR based on the 
\citet{HB06} SFR. Any 
further flattening of the SFR at high redshift would produce a greater abundance of metals and lead to an
excessive optical depth.  Adopting an alternate normalization based on the \citet{Behroozi} SFR leads to 
a lower SFR at high redshift by a factor of approximately 0.3 dex \citep{Trenti,bs}.
One can obtain a more realistic SFR by assuming that two key time-scales are proportional with a fixed constant of proportionality: the inverse specific star formation
and mass accretion rates. This assumption fits all data to $z\sim 8$ and has predictive power to  higher $z\ltsim 15$.
By a small adjustment in the slope of the SFR at high redshift, we can 
obtain an excellent fit to the \citet{Behroozi} normalization of the SFR based on the GRB rate.
This is shown in  Fig.~\ref{fig:k2}, where we now fit the 
five data points (colored magenta) which are scaled down from the five points from \citet{Kistler13} shown in 
Fig.~\ref{fig:k1}a.
As one can see by the black dashed curve in the upper right panel of Fig.~\ref{fig:k1}b, we now obtain a significantly 
better fit for the optical depth, $\tau = 0.088$, though the redshift of reionization remains somewhat low, 
$z_I = 8.96$.
Here we have simply raised the parameter $a$ from 1.92 to 2.0. Note that the SFR rate falls slightly below the central values of the GRB data using the \cite{HB06} normalization and is within the uncertainties of the 
SFR data shown in Fig.~\ref{fig:k1}a. The metallicity and SNII rate in this case is very similar to that
in model 1, as seen by comparing the red solid and dashed curves in the lower panels of Fig.~\ref{fig:k1}c, d.

\subsection{The SFR based on the observations of star-forming galaxies}

The most conservative approach in obtaining the average SFR at high redshift
comes from the direct observation of the galaxy luminosity function which yields the volume density
of galaxies as a function of luminosity. Over the last 10-15 years, the data on the luminosity function
have been extended out to high redshift with the most recent data reaching $z \sim 8- 10$ \citep{O13-2,O14} 
(for a comprehensive discussion of these observational advances, see \citet{B14}). Extracting
the SFR, however, requires some knowledge of the halo mass function and in particular
the ratio of the total stellar mass to halo mass at high redshift. Here we adopt the results of the 
recent analysis in \citet{Behroozi} where semi-analytical models of cosmological volumes were used to perform abundance matching with the galaxy luminosity function data to high $z$ and to extract the halo star formation efficiency. A key conclusion is that stellar mass efficiency increases beyond $z\sim 4,$  otherwise one would have 
too little stellar mass or star formation rate associated with galaxies at $z\sim 8.$ Other studies reach similar conclusions \citep{2010ApJ...714L.202T, 2014MNRAS.439.1326W}.

\begin{figure}[htb!]
\begin{center}
\epsfig{file=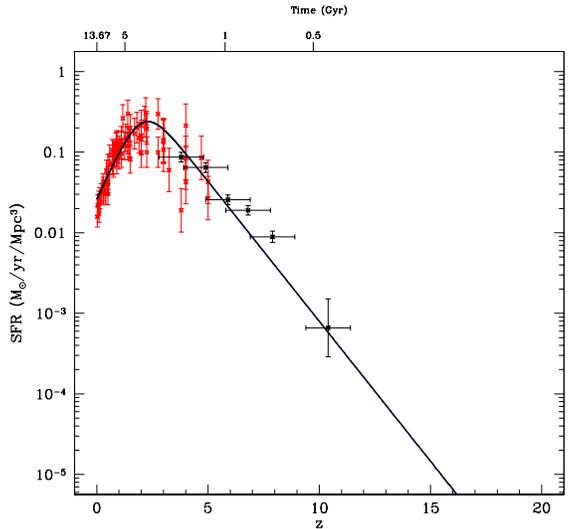, height=3in}
\end{center}
\caption{As in the upper left panel of Fig.~\ref{fig:k1}a using the SFR from \citet{Behroozi} (which includes points from \cite{B14, O13-2} shown in black) from high $z$ observations of the luminosity function (model 3). 
\label{fig:b1}
}
\end{figure}

Fitting the \citet{sp03} form given in Eq. (\ref{shsfr}) to the SFR derived by \citet{Behroozi},
we find  $\nu = 0.24$  M$_{\odot}$/yr/Mpc$^{3}$, 
$z_m = 2.3, a = 2.2$, and $b = 1.4$. Though the parameter values used here
appear to be quite similar to those used in Sect \ref{kist} above, the part of the SFR which 
is of most interest here is that at large redshift, where the SFR goes as $e^{(b-a)z}$ and thus
declines much more rapidly in this case. 
This SFR is shown in Fig.~\ref{fig:b1}. 
As one can see by the solid blue curve in the upper right panel of Fig.~\ref{fig:k1}b,  using only a single mode 
of star formation in this case is not sufficient to account for the optical depth derived from CMB data. 
For this choice of the SFR, the optical depth only rises to $\tau \approx 0.055$ at high redshift, 
more than 2 $\sigma$ below the CMB value. Furthermore, the redshift of reionization is significantly
below the WMAP value. For this case, we find $z_I = 6.41$ and is over 4 $\sigma$ too low.
Nevertheless, the overall metallicity and Type II supernova
rate are quite consistent with observations as seen by the solid blue curves in the lower panels of Fig.~\ref{fig:k1}c, d.
Thus in this case, we are led to consider a high mass mode of star formation operating predominantly at high
redshift, to fit the optical depth.

For the high mass mode, we assume an IMF with a Salpeter slope (as we have done for 
all other IMFs) but with a restricted mass range from 36 -- 100 M$_\odot$.
Although we also use the \citet{sp03} form for the SFR, there is now, in principle, 
considerable freedom for the selection of the four SFR parameters. To help in this choice 
we performed a scan of the 4d parameter space to minimize a $\chi^2$ function which is described below. 
We include several constraints in establishing the $\chi^2$ function:
\begin{itemize}
\item{{\bf The optical depth:} We include two contributions to $\chi^2$ based on
obtaining the correct value for the optical depth at high redshift (here taken to be $z = 30$).
We use the WMAP value of $\tau = 0.089 \pm 0.014$.  We also take in account the redshift of 
reionization, $z_I$, defined as $Q_{\rm ion} = 0.5$. We take $z_I = 11.1 \pm 1.1$. }
\item{{\bf The overall metallicity:} Because of the considerable amount of scatter in the metallicity data
at high redshift, we use an approximate mean of the observed data with a generous uncertainty. 
The main goal is to insure that the computed metallicity is in the right ballpark. 
We compute the overall metallicity, $Z$, at redshifts $z = 4$ and 
$z = 0$.  We have taken the observed values $Z(4) = 0.03 \pm 0.01$ and $Z(0) = 0.5 \pm 0.5$.
Units for $Z$ are taken to be relative to the solar metallicity. }
\item{{\bf Individual element abundances:} The global metallicity is  not sufficient for breaking
inherent degeneracies in the parameter space and therefore we include as constraints the abundances
of  carbon and oxygen compared to the high redshift data taken from the SAGA DataBase, \citep{suda08,suda11}} for roughly 140 objects
for each element. The redshift for each observable is converted within the hierarchical model from
the observed iron abundance.   
\item{{\bf The SFR at high redshift:}  We also include as a constraint the observed SFR
at redshifts $z \approx 6 - 10$ comprising of 5 data points from \citet{O13-2}.}
\end{itemize}
The $\chi^2$ function is then defined by taking the difference between a computed value and
the data (squared) and weighted by an observational uncertainty.

\begin{figure}[htb!]
\begin{center}
\epsfig{file=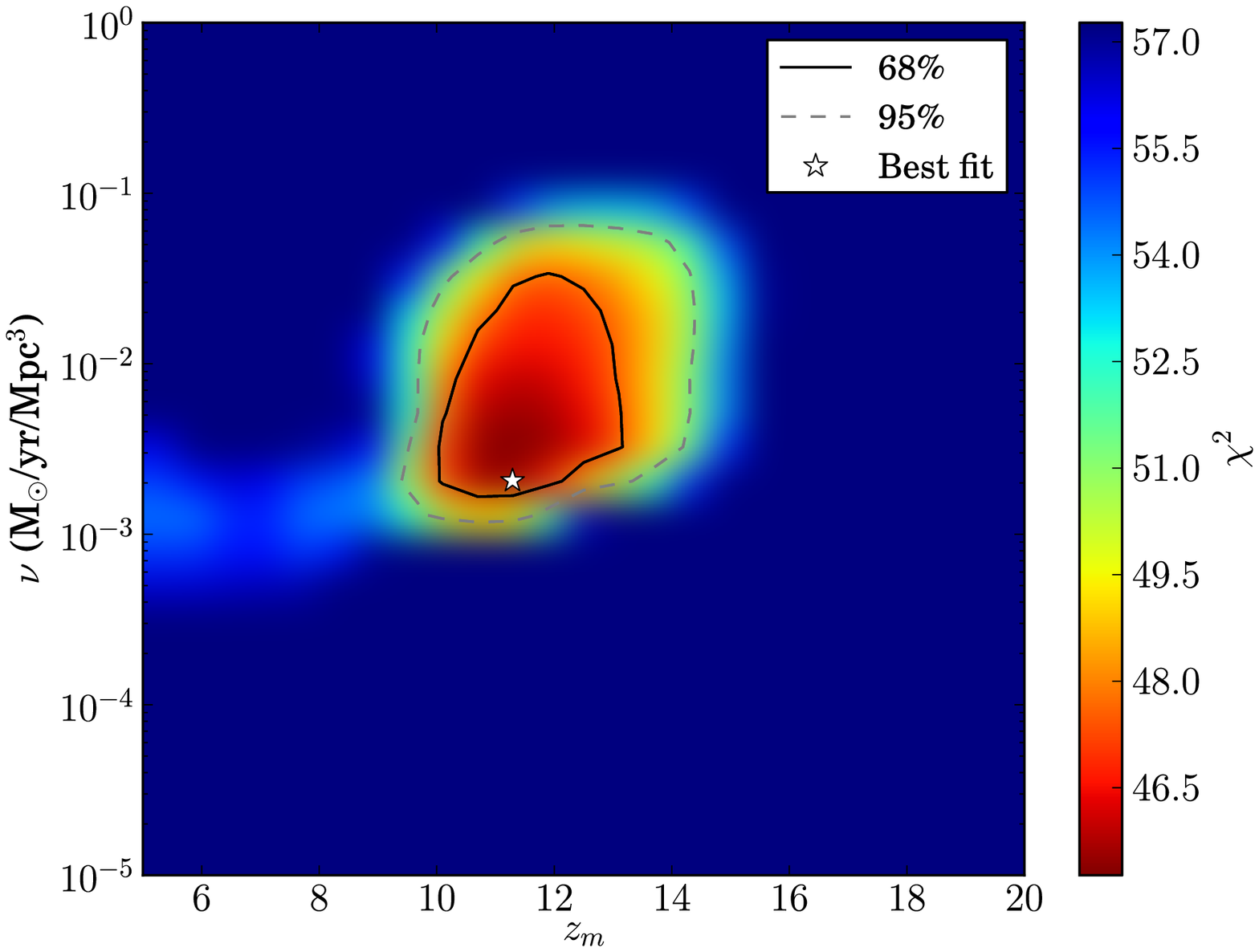, height=3.3in, angle=-0}
\epsfig{file=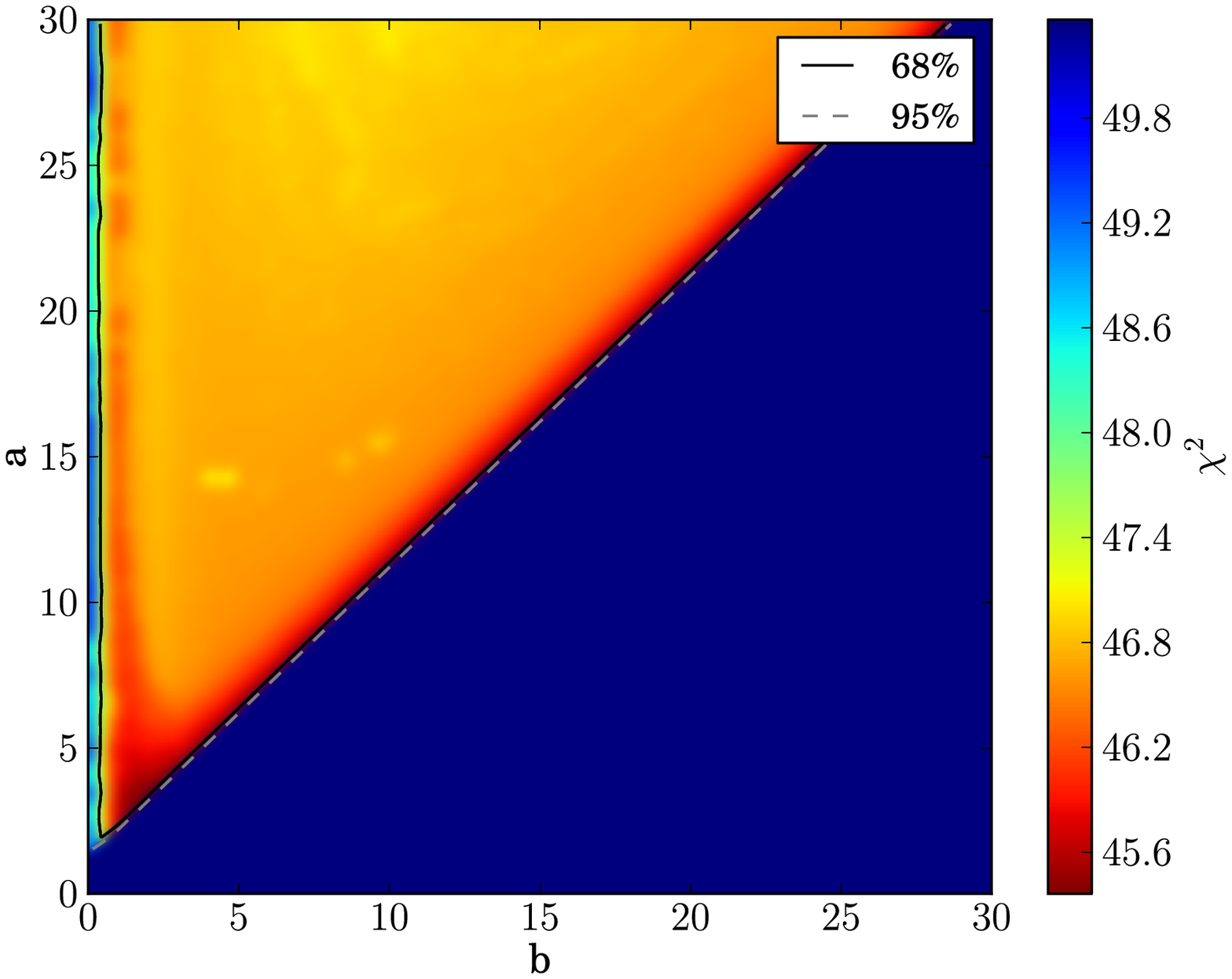, height=3.3in, angle=-0}
\end{center}
\caption{ a)  2D maps of $\chi^2$ in the $\nu, z_m$ parameter plane (upper) and 
$a, b$ parameter plane (lower). 68\% (solid) and 95\% (dashed) frequentist confidence intervals are shown, calculated using $\Delta \chi^2$ values of 2.30 and 5.99, respectively. The location of the best fit point in the
upper panel is shown by a star. In the lower panel, the $\chi^2$  function runs along the diagonal
and changes very little along that diagonal.
\label{fig:like}
}
\end{figure}

Some results of the frequentist likelihood analysis are shown in Fig. \ref{fig:like}
where we show 2D parameter planes in $\nu, z_m$ (left) and 
$a, b$ (right) for the high mass mode parameters. 
In the left panel, we have divided the 2D $\nu$, $z_m$ parameter space into a 100 by 100 grid and minimized the $\chi^2$ in each bin using a Nelder-Mead simplex method \citep{NM}.  $\nu$ and $z_m$ are not strictly fixed, but are allowed to vary within the boundaries of each bin; the other two parameters, $a$ and $b$, are allowed to vary between 0 and 30.  These boundaries have proved sufficient to fully explore the areas of minimal $\chi^2$ within the model parameter space.  The 68\% and 95\% frequentist confidence intervals are calculated about the minimum $\chi^2$ value, using $\Delta \chi^2$ values of 2.30 and 5.99, respectively (based on a $\chi^2$ distribution with two degrees of freedom).  The same procedure is repeated in the 2D $a$, $b$ parameter space, allowing $\nu$ and $z_m$ to vary between $10^{-5}$ and 1 M$_\odot$/yr/Mpc$^3$ and 5 and 20, respectively.

Because the normal mode of star formation is already a good fit to the SFR,
we expect the high mass mode to be rather strongly peaked at high redshift. Due to the 
nature of the above constraints (contributions to $\chi^2$) we can deduce that the 
peak of the high mass mode will occur near $z = 11$ so as to ensure the correct 
epoch of reionization.  The evolution of the individual elements such as C and O
serve as a limiting factor in the astration rate, $\nu$, as there is no direct constraint
available from the SFR itself. 
As one can see in the upper panel of Fig.~\ref{fig:like}, the $\chi^2$ function increases very
rapidly as $\nu$ is decreased below $10^{-3}$M$_\odot$/yr/Mpc$^3$.
At smaller $\nu$, the contribution of the high mass mode becomes insignificant.
The best fit lies slightly above the 68\% contour as is indicated by a star. 
The best fit occurs at $\nu = 0.00287$ M$_{\odot}$/yr/Mpc$^{3}$, $z_m = 11.4$.
As one can see in the lower panel of Fig.~\ref{fig:like},
 the $\chi^2$ likelihood function is extremely flat along the line where $a \ga b$. We have
 taken a central point at $a = 13.2$, and $b = 12.4$.
The large values of the slope parameters $a$ and $b$, lead to a 
sharply peaked SFR for the high mass mode.

\begin{figure}[htb!]
\begin{center}
\epsfig{file=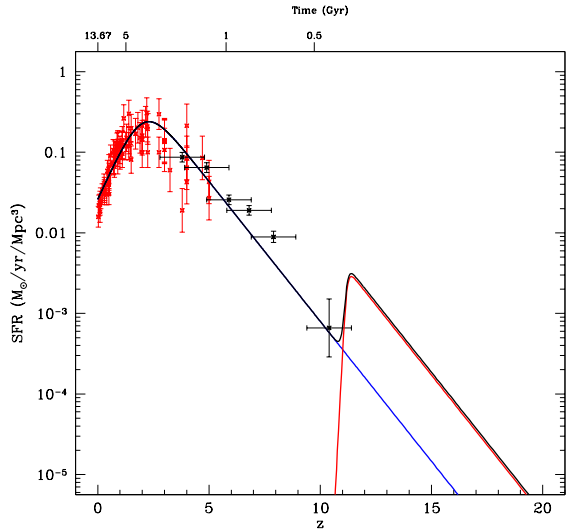, height=3in}
\end{center}
\caption{ As in Fig.~\ref{fig:b1} with the addition of a high mass mode.
\label{fig:b2}
}
\end{figure}

Adopting the best fit value for the parameters, we find the star formation rate shown in 
Fig.~\ref{fig:b2}. The blue curve is the same as in model 3 (shown in Fig.~\ref{fig:b1}) and the red
curve corresponds to the high mass model.  The black is the total SFR. 
As one can see, the high mass mode contributes 
very little to the SFR at $z \lesssim 10$ where data are available and is sharply peaked at 
$z \approx 11$. In the upper right panel of  Fig.~\ref{fig:k1}b, we see a good match to the
optical depth (shown as the blue dashed curve) as a function of redshift. 
In this case $\tau = 0.080$ and $z_I = 11.15$ in almost perfect agreement
with the WMAP value for $z_I$.
At the redshifts shown
in the lower panels of Fig.~\ref{fig:k1}c, d, there is virtually no difference between the metallicity and SNII rates
in models 3 and 4.

\section{Resulting Chemical Evolution}

Having laid out our selection of models for the SFR, we are now in a position 
to explore the consequences for chemical evolution.  In particular we will be interested in the evolution of
the abundances of several elements. In order to compute the element abundances, we 
utilize mass and metallicity-dependent yields taken from the tables of yields  from \citet{ww95}  for massive stars (10$< M/M_{\odot}<$40). An interpolation is made between the different metallicities (Z=0, 0.0001 Z$_\odot$, 0.001 Z$_\odot$, 0.1 Z$_\odot$ and Z$_\odot$ ) and we extrapolate the tabulated values beyond 40 M$_\odot$.

\begin{figure}[htb!]
\begin{center}
\epsfig{file=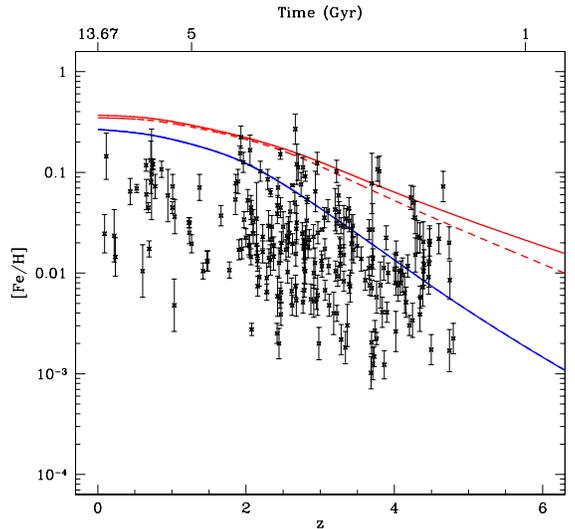, height=3in}
\end{center}
\caption{  The evolution of the iron abundance as a function of redshift for the four models under consideration. Description of data and curves can be found in the text. 
\label{fig:Fe}
}
\end{figure}

Our numerical results can now be compared to various observations for
each element under consideration. Iron (shown in
Fig.~\ref{fig:Fe}) is measured in damped Ly$\alpha$ (DLA) systems as a
function of redshift. As noted earlier, when discussing the overall
metallicity, most of the high $z$ iron data points (47) come from
\cite{rafelski12} and 195 others from previous surveys
\citep{prochaska05, prochaska09}.
Note that the iron abundances measured by \cite{rafelski12}  are measured in the gas phase. It is well known that iron is depleted onto dust grains in the 
interstellar medium. Therefore these measurements should be considered as lower limits.

\begin{figure}[htb!]
\begin{center}
\epsfig{file=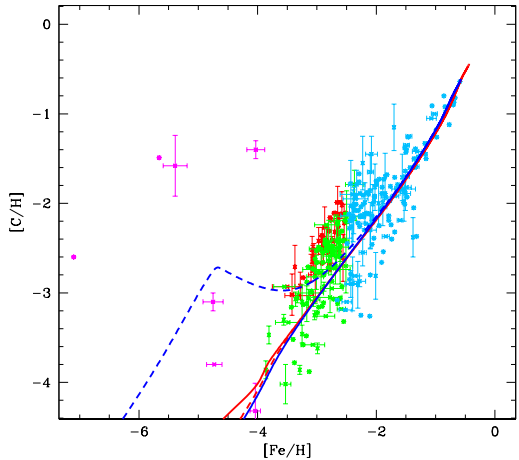, height=2.8in}
\end{center}
\caption{The evolution of the carbon abundance as a function of the iron abundance for the four models under consideration. A description of the data and curves can be found in the text.
\label{fig:C}
}
\end{figure}

\begin{figure}[htb!]
\begin{center}
\epsfig{file=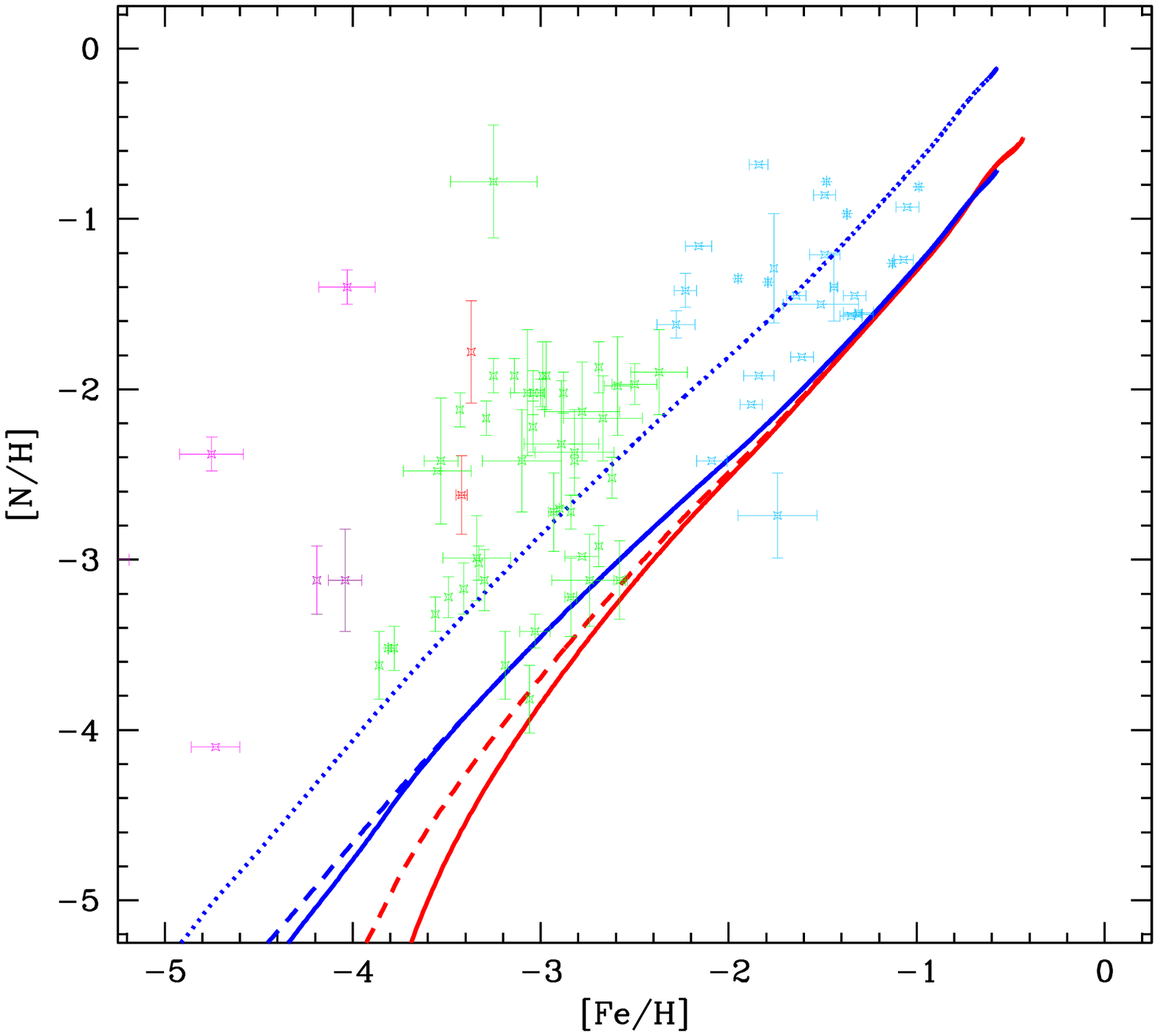, height=2.8in}
\epsfig{file=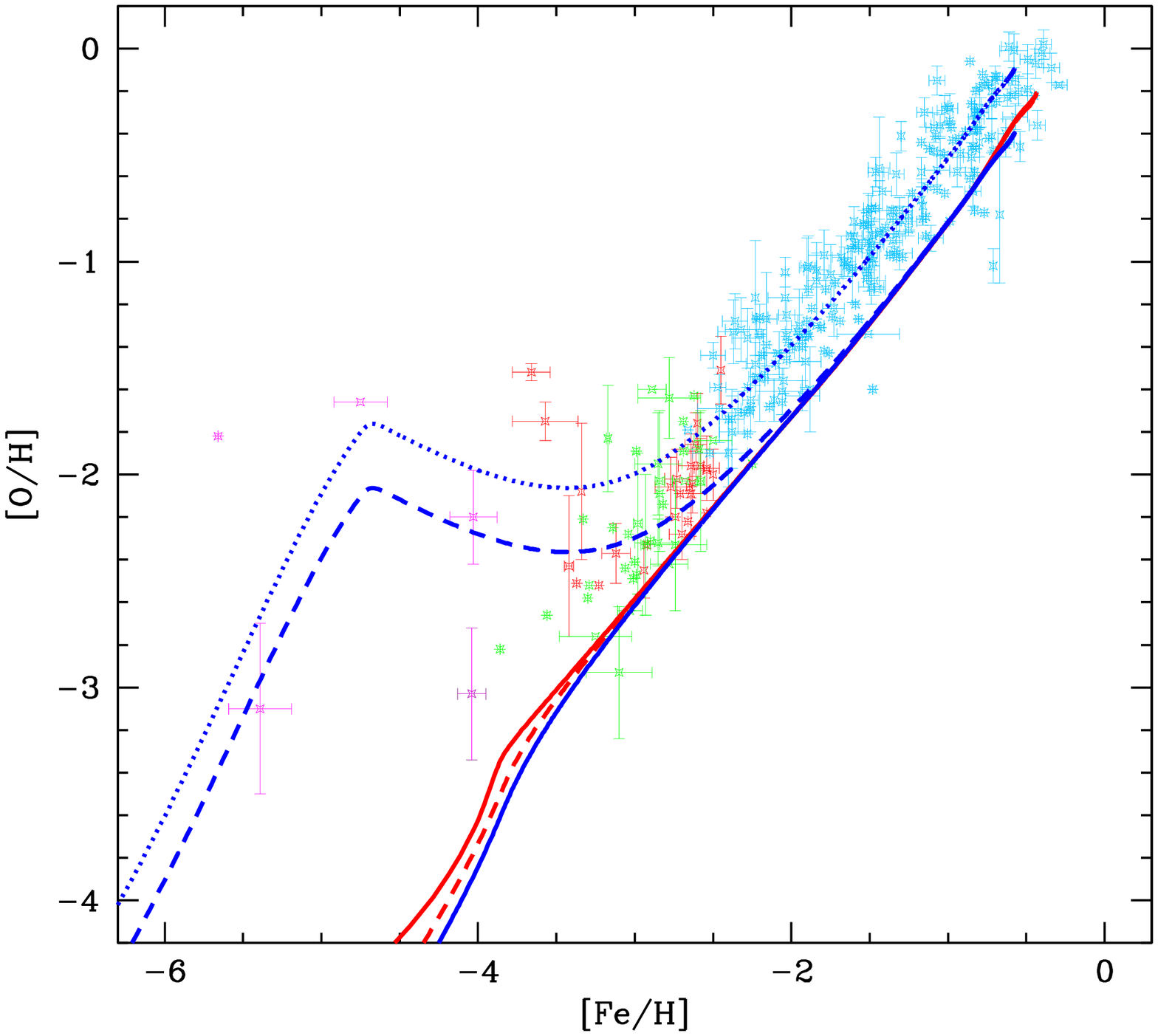, height=2.8in}
\end{center}
\caption{ The evolution of the nitrogen (left) and oxygen (right) abundances as a function of the iron abundance for the four models under consideration. The blue dotted curves show the evolution of N and O
using an alternate set of enhanced yields in the case of model 4.
\label{fig:NO}
}
\end{figure}

\begin{figure}[htb!]
\begin{center}
\epsfig{file=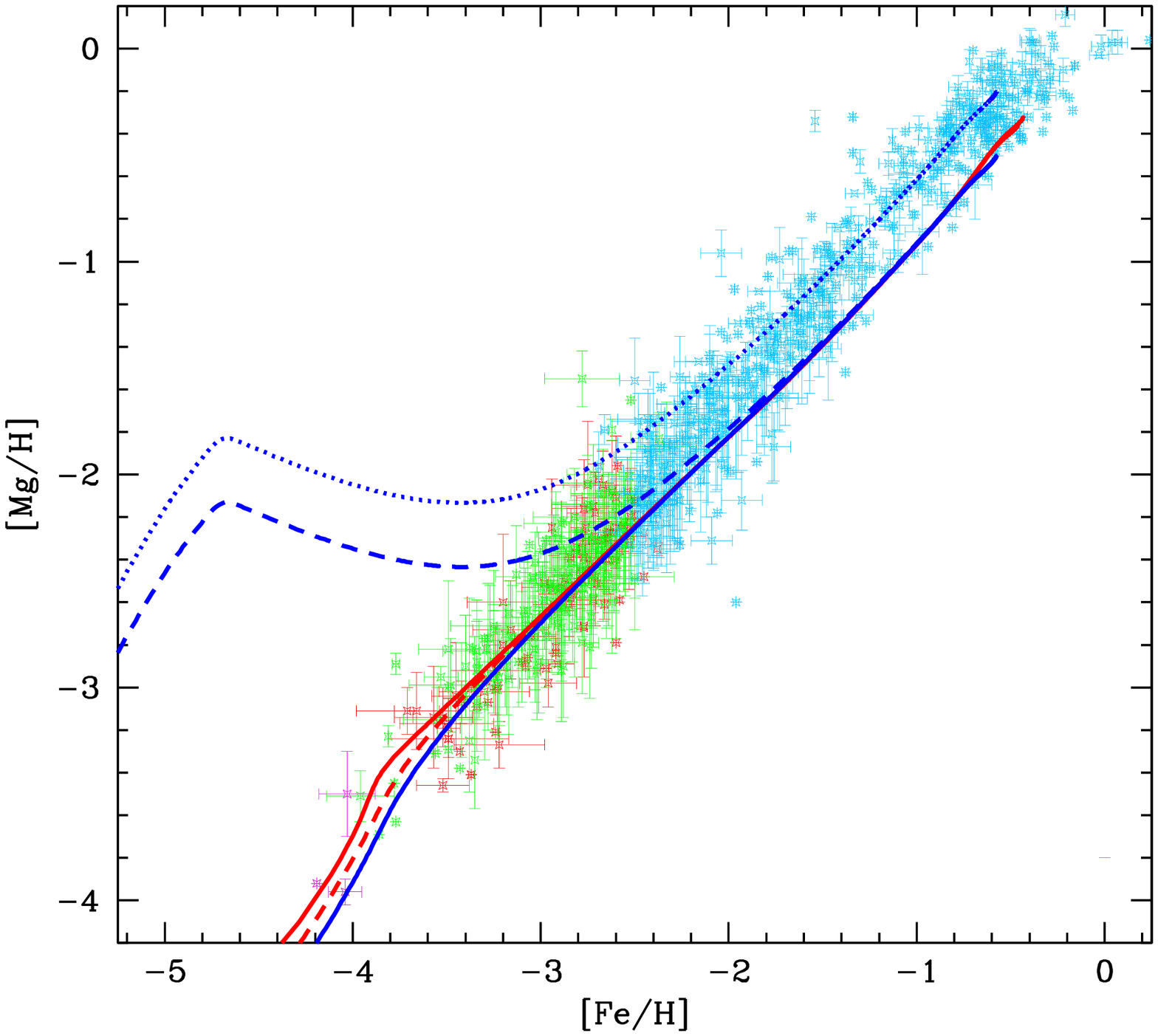, height=2.8in}
\epsfig{file=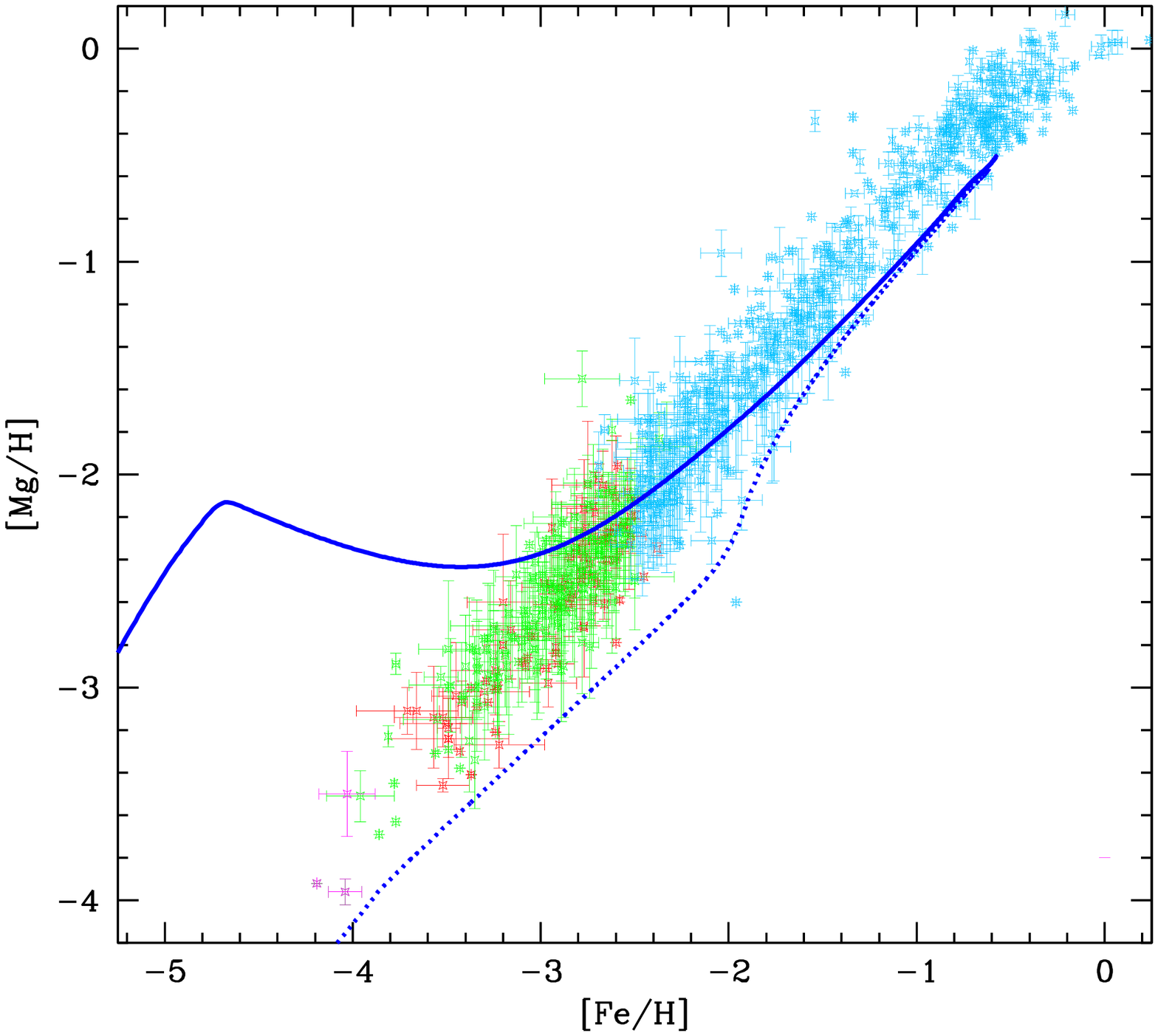, height=2.8in}
\end{center}
\caption{ As in Fig.~\ref{fig:NO}, the evolution of the magnesium abundance as a function of the iron abundance for the four models under consideration. In the left panel, we use a Mg yield coming from \cite{ww95}. The blue dotted curve shows the evolution of Mg
using an alternate enhanced yield in the case of model 4, multiplying  the yield by a factor 2. In the right panel, we use another yield, coming from \cite{hw10} (blue dotted line) compared with the result from model 4 (solid blue line).
\label{fig:Mg}
}
\end{figure}

Concerning the $\alpha$ elements: carbon, oxygen, magnesium  and nitrogen\
(shown in Fig.~\ref{fig:C}, Fig.~\ref{fig:NO} and Fig.~\ref{fig:Mg}),
the data points come from the SAGA DataBase - Stellar Abundances for the
Galactic Archeology \citep{suda08, suda11}. In this compilation we have
selected the bulk of standard observations (blue points corresponding to
dwarfs, red and green ones to metal poor dwarfs and giant stars
respectively and magenta points represent the ultra metal poor stars). 

For each quantity, we will compare the resulting evolution using four
different assumed SFRs based on: 1) the GRB rate from \citet{Kistler13} -
model 1; 2) the renormalized rate from \cite{bs} - model 2; 3) the
observation of star forming galaxies from \cite{Behroozi} without a
high mass mode - model 3; 4) the same with a high mass mode as discussed
above - model 4.

We begin the discussion with the evolution of iron.
In Fig.~\ref{fig:Fe}, we show the evolution of the iron abundance
([Fe/H] corresponds to the log of the ratio of iron H relative to the
solar ratio) as a function of redshift for the 4 models under consideration.
The two upper curves, shown in red, correspond to the SFR based on the
GRB rate (models 1 and 2). Not surprisingly, they show the highest iron
abundance at any redshift. The dashed of the two corresponds to
the lower normalization argued in \citet{bs} (model 2), but the two are
essentially equivalent. The lower two curves, shown in blue (models 3 and
4), correspond to the SFR based on the luminosity function,
with (dashed - model 4) and without (solid - model 3) the high mass
mode. These too, are essentially indistinguishable because the Pop III star
mode is effectively efficient only at very high redshift.
Recall, as noted above, these iron measurements should be considered as lower limits.

While one could argue that the luminosity based SFR provides a better
fit to the data, it must be noted that (i) the mean metallicity
weighted by the HI column density derived by \cite{rafelski12}
lies between the blue and  red curves at any redshift and (ii) the
slope of the observed evolution is closer to that of the GRB based
models so that these models may better represent reality at higher
redshift.

We next consider the evolution of carbon. In Fig.~\ref{fig:C}, we show the evolution [C/H] as a function of the iron abundance  for the 4 models
under consideration. 
As one can see for the figures, the carbon abundance is essentially independent of our 
choice of model for [Fe/H] $> -2$ or $z < 4$.  At higher $z$ (lower metallicity), model 4 actually does a 
better job at an explanation of the ultra metal poor  carbon enhanced metal poor  (CEMPs) stars, though as one can see there is considerable dispersion in these points ( shown in magenta). We have not considered these in detail here as a specific study including an intermediate mass stellar mode would be required. Thus, in this context, the carbon evolution we display is only a lower limit at very low metallicity.

In Fig.~\ref{fig:NO}, we show the analogous evolution of nitrogen and
oxygen as a function of [Fe/H].
The nitrogen abundances are very similar for all four models considered.
They all slightly underproduce [N/H]. Note however, that our predictions
are closer to the measurements in damped Lyman-$\alpha$ systems
(\citep{petitjean08, pettini08, zafar14}
for which [N/O]~$\sim$~$-1$ for [O/H] in the range between $-1$ and $-3$.

Nitrogen has different nucleosynthetic origins (compared to oxygen for example), including one from a
potential intermediate mass stellar mode which is not considered here.
Consequently, our calculation provides only a lower limit to the
production of nitrogen. To have an idea of how much additional N
production is necessary to match the data, we have multiplied the N
yield by a factor of 4 and show the result for model 4 by the dotted
blue curve.
In contrast to C and O, N is not produced in massive stars, so model 4
does not show a bump at low metallicity.

The oxygen abundances for models 1, 2, and 3 all look similar, however, the evolution 
of [O/H] is significantly different for model 4. 
Here we see directly the impact of the high mass mode
which shows a local peak in [O/H] at [Fe/H] $\approx -4.8$. This is the only model that can 
explain the abundances seen in very low metallicity  stars (magenta points), as is the case for carbon as well.  Due to uncertainties concerning the oxygen synthesis in massive stars, we show the 
[O/H] abundance for model 4 using an O yield enhancement of a factor of 2 (shown by the dotted blue curve). 
The result for the other models
would scale similarly and provide a better fit to the data at [Fe/H] $> -2$.

As in the case for oxygen, the evolution of magnesium shows the impact of the 
high mass mode in model 4 as seen in Fig.~\ref{fig:Mg} (left) where the evolution of [Mg/H]
is shown as a function of the iron abundance. 
 While models 1,2 and 3 all fit the data nicely, model 4 appears to overproduce 
Mg at [Fe/H] $< - 3$. Once again, for comparison, we have plotted (dotted blue line) the Mg evolution multiplying the Mg yield by a factor 2.  A recent study \citep{hw10} devoted to nucleosynthesis in massive stars at zero metallicity presents new Mg yields.  There, it was found that no Mg is produced for massive stars larger than 30 M$_\odot$ in the most of the models considered at zero metallicity. We have included these new yields at very low metallicity (Z$<$ 0.0001 Z$_\odot$) which should result in the lowest possible Mg production.
The right panel presents the Mg evolution for model 4 (which includes a massive mode at high redshift) with these two different yields: from \cite{ww95} (solid line) and from \cite{hw10} (dotted line). This choice of yields
for the early production of Mg provides us with a lower limit to Mg/H and fits the lower envelope of data.

A signature of a high mass mode which corresponds to a distinct population of stars (Pop III)
is clearly of great importance in understanding the chemical evolutionary history of the universe.
\citet{frebel07} defined a  transition discriminant, \Dtrans, based on the carbon and oxygen abundances
at low metallicity,  and recently \cite{frebel13} have provided an updated formula for \Dtrans:
\begin{equation}
\Dtrans\equiv \log_{10}(10^{\rm [C/H]}+0.9\times10^{\rm [O/H]})\, ,
\end{equation}
which we use here.
When
sufficiently abundant, ionized carbon and neutral atomic oxygen  
act as a trigger to lower mass star formation and signify the transition to Pop I/II star formation 
\citep{bromm:03,yoshida:04}. For a more detailed discussion of this quantity in the context
of chemical evolution with hierarchical structure formation, see \citet{rollinde}.

\begin{figure}[htb!]
\begin{center}
\epsfig{file=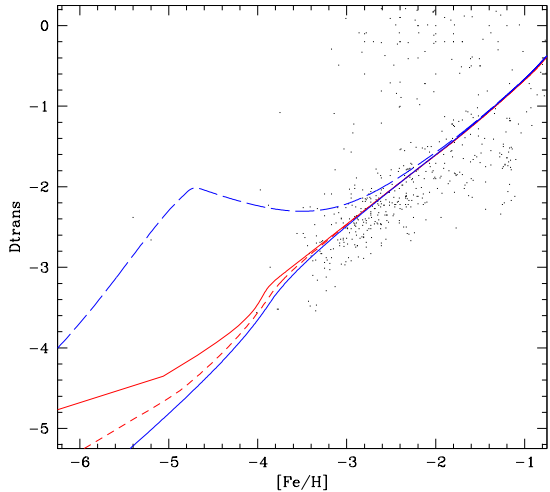, height=3in}
\end{center}
\caption{The evolution of the $\Dtrans$  as a function of the iron abundance 
for the four models under consideration.
\label{fig:dtrans}
}
\end{figure}

In Fig.~\ref{fig:dtrans}, we show the evolution of $\Dtrans$  as a function of [Fe/H] for the four models.
All models make similar predictions for [Fe/H] $\ga -3$. Only model 4 can explain
the 2 observations with relatively high \Dtrans at [Fe/H] $\la -5$, while none of the models
can account for the scattering of points with \Dtrans $\ga -1$ and $-3 < $ [Fe/H] $< -1$. As noted earlier, we do not consider here 
the effects of an intermediate mass mode which would be a large contributor of carbon at low metallicity. These points correspond precisely to the high C content in CEMP stars. 

\begin{figure}[htb!]
\begin{center}
\epsfig{file=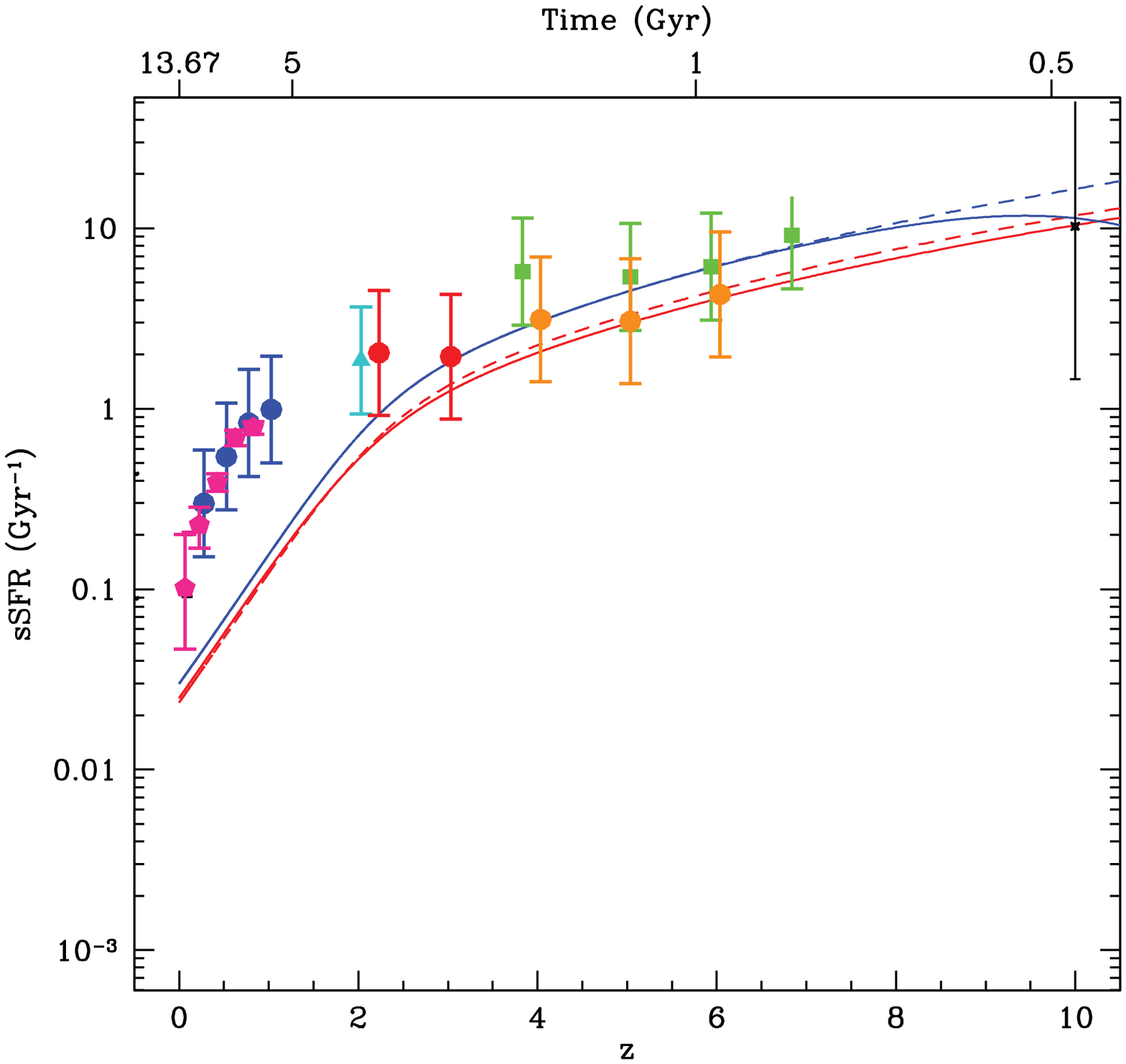, height=3.15in}
\epsfig{file=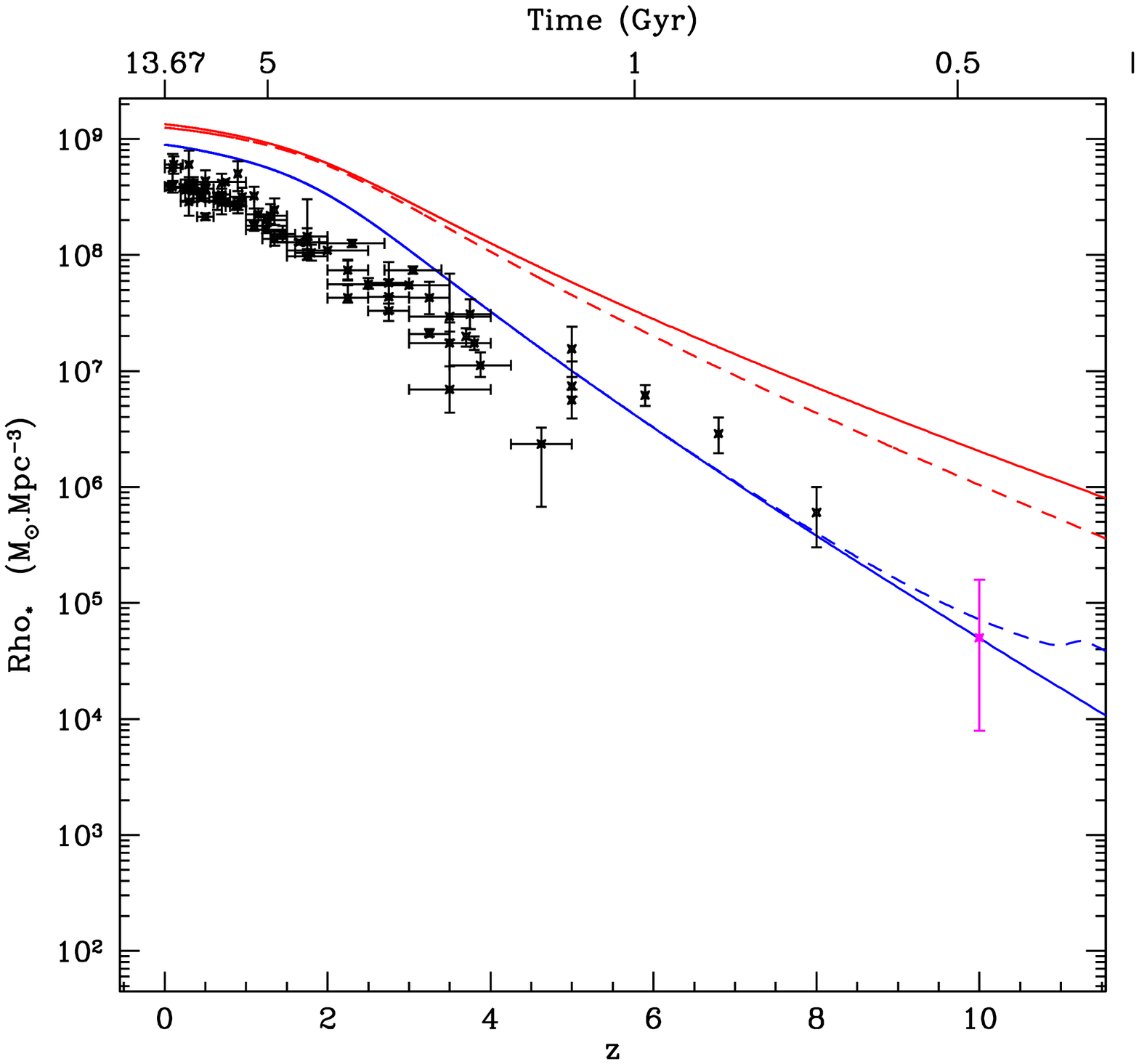, height=3.2in}
\end{center}
\caption{ a) The evolution of the mean specific star formation (sSFR) as a function of the redshift for the four models under consideration. Stellar mass of galaxies are in the range of $10^{9.4}$  to $10^{10}$ M$_\odot$.  Data points come from the literature \citep{daddi07, noeske07, damen09, reddy12, gonzales12, stark13} and have been compiled in \cite{madau14}. The $z=10$ data point comes from \cite{O13-2}. b) The evolution of  stellar mass density as a function of the redshift for the four models under consideration is also shown. The data points are taken from a compilation made by \cite{madau14} (their table 2) using UV and IR data. The magenta $z=10$ data point comes from \cite{O13-2}.
\label{fig:ssfr}
}
\end{figure}

Finally, Fig.~\ref{fig:ssfr} (left) shows the evolution of the mean specific star formation (sSFR) as a function of  redshift for the four considered models. The sSFR is calculated by dividing the cosmic SFR density by the mean stellar mass density (shown in the right panel of Fig.~\ref{fig:ssfr}). We have taken the compilation of data from \citet{madau14} and added a point from  \citet{O13-2} at $z=10$. The estimated stellar mass range is $10^{9.4}$  to $10^{10}$ M$_\odot$. There no significant difference between the models. At high redshift, they fit  the data points.

As usual, at low redshift, the sSFR calculation is too steep relative to data, possibly due to an overestimate of the stellar mass.
The model rise from the present epoch to the $z\sim 2$ star formation rate peak is shallower than found in the data. It is likely that this  is due to hidden star formation in dense molecular gas  and the galaxy merger contribution,  not included in our simple modeling and reflected in the strong difference between the star formation rate contributions by normal star-forming galaxies (relatively shallow and with low sSFR) and luminous IR galaxies (LIRGS) (very steep and high sSFR) as found by \cite{leborgne09}.

The right panel shows the evolution of the mean stellar mass density as a function of redshift. The data points are taken from the compilation  by \citet{madau14}, once again adding the $z=10$  point from \citet{O13-2}. Models 3 and 4 fit the  observational data at high redshift.
 Indeed, these results clearly show the limitation of this approach at low redshift. This is related to feedback processes that prevent stars from forming and enhancing the sSFR. 
Models 1 and 2 cannot fit this data because these latter observations are deduced from the SFR data from observations of star-forming galaxies. 
While GRBs are associated with dwarf galaxies, however
even just considering the data based on the GRB rate, additional  dwarf galaxies are still required that are well below the observational limits. Possibly these might be associated with GRBs that have no apparent host galaxy.

\section{Discussion}

The DLA abundances reflect enhanced star formation rates at high redshift as does the intercluster medium 
at low redshift. The fact that these data are consistent  over a wide redshift range argues 
against a systematic change in the IMF, but favors a dwarf galaxy population that has a steep luminosity function and hosts the GRBs. Evidence for a steep luminosity function comes both locally from  dwarf galaxy surveys in clusters \citep{2006A&A...445...29P}, where harassment is thought to be responsible, 
and at  high redshift where studies of the stellar mass function show that the faint-end slope, $\alpha$, steepens from $\sim 1.6$  to $\sim 2.0$ over $z\sim 4-10$ \citep{B14,2014arXiv1408.2527D},  the explanation in this case  presumably being due to the expected convergence to the halo mass function. The steep value of $\alpha$ has been found via  a lensed sample of high redshift ($z\sim 7-8$) galaxies to extend to below  $0.1 L^\ast $ \citep{2014arXiv1409.0512A}.

WMAP and Planck data \citep{wmap,planck} have given us unprecedented precision
in determining cosmological observables. Among these are the integrated optical depth and the corresponding
epoch of reionization at $z \approx 11$.  We therefore know that some stellar activity and hence 
chemical evolution must have taken place at a still higher redshift (assuming that 
stellar light is the source of reionization). The discovery of Lyman break galaxies at 
$z>3$ \citep{lbreak} opened the door to a host of rest frame UV observations of galaxies at ever increasing
redshift. There are now six candidate galaxies at $z\sim 10$ \citep{B14}, and between 200-500 galaxies at redshift $z\sim 7 - 8$ \citep{B11,fink,mcl} allowing one to map out the global UV luminosity function
and gain insight into the global SFR density. However because observations at $z \sim 10$ and above 
remain sparse, significant uncertainties remain  in the SFR at high redshift. To some extent,
the SFR determined from the galaxy luminosity function can be thought of as setting  a lower limit
as systematic effects due to dust obscuration would tend to increase the derived SFR.
There is also the problem that such surveys are flux-limited and hence biased by the 
brightest galaxies, and again the true SFR might be higher, particularly at high redshift 
\citep{2010ApJ...714L.202T,2014MNRAS.439.1326W}. Indeed, as we have shown above 
(in our discussion of model 3), the SFR derived from these flux-limited surveys
is not sufficient to reionize the universe at sufficiently high redshift. A similar conclusion was reached
in \citet{robetal}.

While the true SFR may be somewhat higher at large $z$ than that derived by the UV galaxy
luminosity function, it is also possible that there was a burst of star formation at $z \approx 11$ 
that was primarily responsible for the reionization of the Universe. Here we have considered
in model 4, the effect of such a burst on the chemical history of the Universe.

On the other hand, GRBs are expected to be visible out to very high redshifts of $z \sim 15 - 20$
\citep{lamb}, the current record being $z\sim 8.2$ \citep{tanvir}. As the origin of GRBs is expected to 
be core collapse supernovae \citep{wb06}, it is sensible to infer that these events may trace the
SFR as many authors have assumed. However, as discussed above, a straight fit to the SFR derived from
existing GRB data would lead to an optical depth in excess of that determined from CMB data. Extracting 
the SFR from GRB is also not free from its own set of uncertainties and biases \citep{Trenti}. Models 
attempting to overcome these problems predict a steeper fall off at high redshift \citep{Trenti,bs}
and these ``softened" SFRs yield predictions for the optical depth and reionization which 
are quite consistent with data as we have shown in model 2. 

Presently,  most observations of chemical abundances are available only at relatively low redshift.
For redshifts $z \ga 5$, we would require observations of objects with iron abundances
[Fe/H] $\la -3$ (see Fig.~\ref{fig:Fe}). Furthermore, in the hierarchical picture of structure formation
which is at the core of our cosmic chemical evolution models, element abundances are 
primarily sensitive to the late-time SFR where the models are constrained by the low
$z$ determination of the SFR. 
As a consequence, in most cases, the models considered
predict very similar abundances for [Fe/H] $\ga -3$.
Differences begin to occur at lower metallicity, and these are most apparent
for model 4 where a high mass mode of star formation is included at high redshift. Most notably
this model predicts enhanced abundances of C, O, and Mg at very low metallicity as is seen in
some of the most primitive stars in our own Galaxy.

It is clear that enhancement of the SFR occurs relative to that inferred from the galaxy luminosity function.
This may be because faint but star-forming galaxies have been missed (as must be supposed if
the GRB data is a better indicator of the true SFR), or because Pop III star formation included
at high redshift results in an exclusively high mass (and short-lived) mode of star formation. At lower redshifts, the SFRs in all models 
are similar (as there is far less uncertainty) and chemical abundances are far less discriminating.
To better disentangle the SFR at high redshift, new abundance data is required at very low metallicity.

\section*{Acknowledgements}

This work has been carried out at  the ILP LABEX (under reference ANR-10-LABX-63) supported by French state funds managed by the ANR 
within the Investissements d'Avenir programme under reference ANR-11-IDEX-0004-02. It was also sponsored by the French Agence Nationale pour la Recherche (A.N.R.) via the grant VACOUL (ANR-2010-Blan-0510-01). EV warmly thanks Damien Le Borgne for continuing and  fruitful discussions. We also thank Yohann Dubois and Christophe Pichon for their friendly help. The work of K.A.O. was supported in part
by DOE grant DE-SC0011842 at the University of Minnesota. 
The work of TP and VM was supported in part by the NSF grant 1204944 at the University of Minnesota.
The research of JS has been supported at IAP by the ERC project  267117 (DARK) hosted by Universit\'e Pierre et Marie Curie - Paris 6, PI J. Silk. JS also acknowledges the support of the JHU by NSF grant OIA-1124403,

\end{document}